\documentstyle[12pt]{article}
\begin{document}
\baselineskip 24pt
\begin{center}
\baselineskip 24pt
\noindent
{\large\bf Running Coupling Constants of Fermions with Masses in Quantum
Electro Dynamics and Quantum Chromo Dynamics}

\vskip 24pt
\normalsize
\baselineskip 24pt
Guang-jiong Ni$^*$\footnote{$^*$ E-mail: gjni@fudan.ac.cn}, Guo-hong Yang,
Rong-tang Fu\\
{\it Department of Physics, Fudan University, Shanghai, 200433, China}\\
Haibin Wang\\
{\it Randall Lab, University of Michigan, Ann Arbor, MI 48109-1120, USA}
\end{center}

\vskip 48pt
\baselineskip 24pt
\normalsize
\begin{center}
{\bf Abstract}
\end{center}

\leftskip 9mm
\rightskip 9mm

\vskip 12pt

\baselineskip 24pt
\indent Based on a simple but effective regularization-renormalization
\baselineskip 24pt
method (RRM), the running coupling constants (RCC) of fermions with masses in
quantum electrodynamics (QED) and quantum chromodynamics (QCD) are calculated
by renormalization group equation (RGE). Starting at
$Q=0$ ($Q$ being the momentum transfer), the RCC in QED
increases with the increase of $Q$ whereas the RCCs for
different flavors of quarks with masses in QCD are different and they increase
with the decrease of $Q$ to reach a maximum at low $Q$ for each
flavor of quark and then decreases to zero at $Q\longrightarrow 0$. The
physical explanation is given.


\newpage
\leftskip 0mm
\rightskip 0mm
\baselineskip 24pt
\normalsize
\noindent
{\bf I. Introduction}

\vskip 24pt
\baselineskip 24pt
\normalsize
\indent  In most literatures and textbooks, the running coupling constant
\baselineskip 24pt
(RCC) in quantum electrodynamics (QED) is usually given as (see, e.g. Ref. [1]
and Eq. (48) below):
\begin{equation}
\alpha(Q)=\frac{\alpha}{1-\frac{2\alpha}{3\pi}\ln\frac{Q}{m_e}}
\end{equation}
where $m_e$ is the electron mass and $\alpha=\frac{e^2}{4\pi}$. Eq. (1) is
very important in physics for it unveils the monotonically enhancing behavior
of electromagnetic coupling constant $\alpha$ in accompanying the increase of
momentum transfer $Q$ between two charged particles and shows the existence of
Landau singularity at an extremely large $Q$. However, in our opinion, there
are still three aspects that can be improved in this paper. (a) Besides
electron, the contributions of other charged leptons and quarks can not be
neglected. (b) While Eq. (1) is scale invariant, it ignores totally the
particle mass effect which is also important at low $Q$ region. (c) While the
normalization in Eq. (1) is inevitably made at
$\alpha(Q=m_e)=\alpha=(137.03599)^{-1}$, we prefer to renormalize it at the
Thomson limit $(Q\longrightarrow 0)$ irrespective of the particle mass.\\
\indent As for quantum chromodynamics (QCD), similarly, the RCC of quark is
usually expressed for
massless quark and so is independent of the flavor of quark. For instance,
it reads [2] (We use the Bjorken-Drell metric throughout this
paper.):
\begin{equation}
\alpha_s(Q)=\frac{4\pi}{\beta_0\ln(Q^2/\Lambda_{QCD}^2)}
\end{equation}
where $\beta_0=\frac{11}{3}C_A-\frac{2}{3}n_f$ with $C_A=3$ and $n_f$
the number of flavors of quarks. The singularity of $Q$ in
Eq. (2), $\Lambda_{QCD}$, is an energy scale characterizing the confinement of
quarks in QCD, $\Lambda_{QCD}\simeq 200MeV$ experimentally.\\
\indent While Eq. (2) successfully shows the asymptotic freedom of quarks at
high $Q$, it is not so satisfying at low $Q$ region, especially for heavy
quarks. The mass of $c$ or $b$ quarks, let alone $t$ quark, is much higher
than $\Lambda_{QCD}$. In other words, the $c$ (or $b$, or $t$) quark does not
exist at low $Q$ regin beneath the threshold for creating $c\bar{c}$ (or
$b\bar{b}$, or $t\bar{t}$) pair and the latter is different for different
flavor. Therefore, instead of Eq. (2), we need a new calculation of
renormalization group equation (RGE) for RCC to discriminate different flavors
of quarks. Evidently, it is necessary to take the mass of quark into account.\\
\indent In recent years, based on the so-called derivative renormalization
method in the literature [3-11], proposed by Ji-feng Yang [12], a simple but
effective renormalization-regularization method (RRM) was used by Ni
{\it et al} [13-17]. It is characterized as
follows. When encountering a superficially divergent Feynman diagram integral
(FDI) at one-loop level, we
first differentiate it with respect to external momentum or mass parameter
enough times until it becomes convergent. After performing integration with
respect to internal momentum, we reintegrate it with respect to the parameter
the same times to return to original FDI. Then instead of divergence, some
arbitrary constants $C_i$ $(i=1,2,\cdots)$ appear in FDI, showing the lack of
knowledge about the model at quantum field theory (QFT) level under consideration. They can only be
fixed by experiments or by some other deep reasons in theory. Since all
constants are fixed at one-loop level, all previous steps can be repeated at
next loop expansion. The new RRM has got rid of
the explicit divergence, the counterterm, the bare parameter and the ambiguous
(arbitrary) running mass scale $\mu$ quite naturally.
In section II we will explain this method by calculating the
RCC in QED [16,18] which also serves as the basis of the following sections. Then
in Sec. III the relevant formulation of RGE for RCC in QCD is presented. The
numerical results are given at Sec. IV. The final section V will contain a
summary and discussion.

\vskip 48pt
\normalsize
\noindent
{\bf II. RGE calculation of RCC in QED}

\vskip 24pt
\indent As is well known, there are three kinds of Feynman diagram integral
\baselineskip 24pt
(FDI) at one-loop level in QED.

\vskip 24pt
\noindent
{\bf 1. Self-energy of electron with momentum $p$}\\
\baselineskip 24pt
\indent The FDI for self-energy of electron reads ($e<0$) [19-22]
\begin{eqnarray}
-i\Sigma(p) & = & (-ie)^2\int\frac{d^4k}{(2\pi)^4}\frac{g_{\mu\nu}}{ik^2}
\gamma^\mu\frac{i}{\not{p}-\not{k}-m}\gamma^\nu \nonumber \\
 & = & -e^2\int\frac{d^4k}{(2\pi)^4}\frac{N}{D}
\end{eqnarray}
$$
\frac{1}{D}=\frac{1}{k^2[(p-k)^2-m^2]}
=\int^1_0\frac{dx}{[k^2-2p\cdot kx+(p^2-m^2)x]^2}
$$
$$
N=g_{\mu\nu}\gamma^\mu(\not{p}-\not{k}+m)\gamma^\nu
=-2(\not{p}-\not{k})+4m.
$$
We first perform a shift in momentum integration: $k\longrightarrow K=k-xp$,
so that
\begin{equation}
-i\Sigma(p)=-e^{2}\int_{0}^{1}dx[-2(1-x)\not{p}+4m]I
\end{equation}
and concentrate on the logarithmically divergent integral
\begin{equation}
I=\int\frac{d^{4}K}{(2\pi)^{4}}\frac{1}{[K^{2}-M^{2}]^{2}}
\end{equation}
with
$$
M^{2}=p^{2}x^{2}+(m^{2}-p^{2})x.
$$
A differentiation with respect to $M^2$ is enough to get
\begin{equation}
\frac{\partial{I}}{\partial M^{2}}=\frac{-i}{(4\pi)^{2}}\frac{1}{M^{2}}.
\end{equation}
Thus
\begin{equation}
I=\frac{-i}{(4\pi)^{2}}[\ln M^{2}+C_{1}]=\frac{-i}{(4\pi)^{2}}\ln
\frac{M^{2}}{\mu^{2}_{2}}
\end{equation}
carries an arbitrary contant $C_{1}=-\ln\mu_{2}^{2}$. After integration
with respect to the Feynman parameter $x$, one obtains
\begin{eqnarray}
\Sigma(p)&=& A+B\not{p} \nonumber \\
A &=& \frac{\alpha}{\pi}m[2-\ln
\frac{m^{2}}{\mu^{2}_{2}}+\frac{(m^{2}-p^{2})}{p^{2}}\ln
\frac{(m^{2}-p^{2})}{m^{2}}] \nonumber \\
B &=& \frac{\alpha}{4\pi}\{\ln
\frac{m^{2}}{\mu^{2}_{2}}-3-\frac{(m^{2}-p^{2})}{p^{2}}
[1+\frac{m^{2}+p^{2}}{p^{2}}\ln\frac{(m^{2}-p^{2})}{m^{2}}]\}.
\end{eqnarray}
Using the chain approximation, one can derive the modification of
electron propagator as
\begin{equation}
\frac{i}{\not{p}-m}\longrightarrow \frac{i}{\not{p}-m}\frac{1}{1
-\frac{\Sigma(p)}{\not{p}-m}}=
\frac{iZ_{2}}{\not{p}-m_{\rm R}}
\end{equation}
\begin{equation}
Z_{2}=(1-B)^{-1}\simeq 1+B
\end{equation}
$$
m_{\rm R}=\frac{m+A}{1-B}\simeq(m+A)(1+B)\simeq m+\delta m
$$
\begin{equation}
\delta m\simeq A+mB.
\end{equation}
For a free electron, the mass shell condition $p^{2}=m^{2}$ leads to
$$
\delta m=\frac{\alpha m}{4\pi}(5-3\ln\frac{m^{2}}{\mu_{2}^{2}}).
$$
We want the parameter $m$ in the Lagrangian still being explained as the
observed mass, i.e., $m_{\rm R}=m_{\rm obs}=m$. So $\delta {m}=0$ leads
to $\ln\frac{m^2}{\mu_{2}^{2}}=\frac{5}{3}$, which in turn fixes the
renormalization factor for wave function
\begin{equation}
Z_{2}=1-\frac{\alpha}{3\pi}.
\end{equation}

\vskip 24pt
\noindent
{\bf 2. Photon self-energy --- vacuum polarization}\\
\baselineskip 24pt
\begin{equation}
\Pi_{\mu\nu}(q)=-(-ie)^{2}{\rm Tr}\int\frac{d^{4}k}{(2\pi)^{4}}
\gamma_{\mu}\frac{i}{\not{k}-m}\gamma_{\nu}\frac{i}{\not{k}-\not{q}-m}.
\end{equation}
Introducing the Feynman parameter $x$ as before and performing a shift
in momentum integration: $k\rightarrow K=k-xq$, we get
\begin{equation}
\Pi_{\mu\nu}(q)=-4e^{2}\int_{0}^{1}dx(I_{1}+I_{2})
\end{equation}
where
\begin{equation}
I_{1}=\int\frac{d^{4}K}{(2\pi)^{4}}\frac{2K_{\mu}K_{\nu}-g_{\mu\nu}K^{2}}
{(K^{2}-M^{2})^{2}}
\end{equation}
with
\begin{equation}
M^{2}=m^{2}+q^{2}(x^{2}-x)
\end{equation}
is quadratically divergent while
\begin{equation}
I_{2}=\int\frac{d^{4}K}{(2\pi)^{4}}\frac{(x^{2}-x)(2q_{\mu}q_{\nu}-
g_{\mu\nu}q^{2})+m^{2}g_{\mu\nu}}{(K^{2}-M^{2})^{2}}
\end{equation}
is only logarithmically divergent like that in Eqs. (5)---(7).
An elegant way for handling $I_1$ is modifying $M^2$ into
\begin{equation}
M^{2}(\sigma)=m^{2}+q^{2}(x^{2}-x)+\sigma
\end{equation}
and differentiating $I_1$ with respect to $\sigma$
two times. After integration
with respect to $K$, we reintegrate it with respect to $\sigma$ two
times, arriving at the limit $\sigma\rightarrow 0$:
\begin{equation}
I_{1}=\frac{ig_{\mu\nu}}{(4\pi)^{2}}\{[m^{2}+q^{2}(x^{2}-x)]
\ln\frac{m^{2}+q^{2}(x^{2}-x)}{\mu_{3}^{2}}+C_{2}\}
\end{equation}
with two arbitrary constants: $C_{1}=-\ln\mu_{3}^{2}$ and $C_{2}$.
Combining $I_{1}$ and $I_2$ together, we find
\begin{equation}
\Pi_{\mu\nu}(q)=\frac{8ie^{2}}{(4\pi)^{2}}(q_{\mu}q_{\nu}-g_{\mu\nu}q^{2})
\int_{0}^{1}dx(x^{2}-x)\ln\frac{m^{2}+q^{2}(x^{2}-x)}{\mu_{3}^{2}}
-\frac{i4e^{2}}{(4\pi)^{2}}g_{\mu\nu}C_{2}.
\end{equation}
The continuity equation of current induced in the vacuum polarization
[19]
\begin{equation}
q^{\mu}\Pi_{\mu\nu}(q)=0
\end{equation}
is ensured by the factor ($q_{\mu}q_{\nu}-g_{\mu\nu}q^2$). So we set
$C_{2}=0$. Consider the scattering between two electrons via the
exchange of a photon with momentum transfer $q\rightarrow 0$ [19].
Adding the contribution of $\Pi_{\mu\nu}(q)$ to tree diagram amounts to
modify the charge square:
$$
e^{2}\rightarrow e_{\rm R}^{2}=Z_{3}e^{2}
$$
\begin{equation}
Z_{3}=1+\frac{\alpha}{3\pi}(\ln\frac{m^{2}}{\mu_{3}^{2}}-
\frac{q^{2}}{5m^{2}}+\cdots).
\end{equation}
The choice of $\mu_{3}$ will be discussed later. The next term in
expansion when $q\neq 0$ constributes a modification on Coulumb
potential due to vacuum polarization (Uehling potential).

\vskip 24pt
\noindent
{\bf 3. Vertex function in QED}\\
\baselineskip 24pt
\begin{equation}
\Lambda_{\mu}(p',p)=(-ie)^{2}\int\frac{d^{4}k}{(2\pi)^{4}}\frac{-i}{k^{2}}
\gamma_{\nu}\frac{i}{\not{p'}-\not{k}-m}\gamma_{\mu}\frac{i}{\not{p}-\not{k}-m}
\gamma^{\nu}.
\end{equation}
For simplicity, we consider electron being on the mass shell:
$p^{2}={p'}^{2}=m^2$, $p'-p=q$, $p\cdot q=-\frac{q^{2}}{2}$. Introducing the
Feynman parameter $u=x+y$ and $v=x-y$, we perform a shift in momentum
integration:
$$
k\rightarrow K=k-(p+\frac{q}{2})u-\frac{q}{2}v.
$$
Thus
\begin{equation}
\Lambda_{\mu}=-ie^{2}[I_{3}
\gamma_{\mu}+I_{4}]
\end{equation}
\begin{equation}
I_{3}=\int_{0}^{1}du\int_{-u}^{u}dv\int\frac{d^{4}K}{(2\pi)^{4}}
\frac{K^{2}}{(K^{2}-M^{2})^{3}}
\end{equation}
\begin{equation}
M^{2}=(m^{2}-\frac{q^{2}}{4})u^{2}+\frac{q^{2}}{4}v^{2}
\end{equation}
\begin{equation}
I_{4}=\int_{0}^{1}du\int_{-u}^{u}dv\int\frac{d^{4}K}{(2\pi)^{4}}
\frac{A_{\mu}}{(K^{2}-M^{2})^{3}}
\end{equation}
\begin{eqnarray}
A_{\mu}&=&(4-4u-2u^{2})m^{2}\gamma_{\mu}+2i(u^{2}-u)mq^{\nu}\sigma_{\mu\nu}
\nonumber \\
& & -(2-2u+\frac{u^{2}}{2}-\frac{v^{2}}{2})q^{2}\gamma_{\mu}
-(2+2u)vmq_{\mu}
\end{eqnarray}
Set $K^{2}=K^{2}-M^{2}+M^{2}$, then $I_{3}=I_{3}^{'}-\frac{i}{32\pi^{2}}$.
$I_{3}^{'}$ is only logarithmically divergent and can be treated as before
to be
\begin{equation}
I'_{3}=
\frac{-i}{(4\pi)^{2}}\int_{0}^{1}du\int_{-u}^{u}dv
\ln\frac{(m^{2}-\frac{q^{2}}{4})u^{2}+\frac{q^{2}}{4}v^{2}}{\mu_{1}^{2}}
\end{equation}
with $\mu_{1}^{2}$ an arbitrary constant. Now $q^{2}=-Q^{2}<0$
($Q^{2}>0$)
\begin{equation}
I_{3}=\frac{-i}{(4\pi)^{2}}\{\ln\frac{m^{2}}{\mu_{1}^{2}}
-\frac{5}{2}+\frac{1}{\omega}F(\omega)\}
\end{equation}
$$
F(\omega)=\ln \frac{1+\omega}{1-\omega}, \hskip 0.3in
\omega=\frac{1}{\sqrt{\frac{4m^2}{Q^2}+1}}.
$$
On the other hand, though there is no ultra-violet divergence in
$I_{4}$, it does have infrared divergence at $u \rightarrow 0$. 
For handling it, we
introduce  a lower cutoff $\eta$ in the integration with respect
to $u$
\begin{eqnarray}
I_{4}&=&\frac{i}{2(4\pi)^{2}}\{[4\ln\eta+5]\frac{4w}{Q^2}F(w)m^{2}
\gamma_{\mu}+\frac{i4w}{Q^{2}}F(w)mq^{\nu}\sigma_{\mu\nu}  \nonumber \\
& &
+4(2\ln\eta+\frac{7}{4})wF(w)\gamma_{\mu}+[\frac{1}{w}F(w)-2]\gamma_{\mu}\}.
\end{eqnarray}
Combining Eqs. (30) and (31) into Eq. (24), one arrives at
\begin{eqnarray}
\Lambda_{\mu}(p',p)&=&
-\frac{\alpha}{4\pi}\{[\ln\frac{m^{2}}{\mu_{1}^{2}}-\frac{3}{2}
+\frac{1}{2\omega}F(\omega)]
\gamma_{\mu}-(4\ln\eta+5)\frac{2\omega}{Q^2}F(\omega)m^{2}\gamma_{\mu}
\nonumber \\
& & -\frac{i2\omega}{Q^2}F(\omega)mq^{\nu}\sigma_{\mu\nu}-2(2\ln\eta+\frac{7}{4})
\omega F(\omega)\gamma_{\mu}\}.
\end{eqnarray}
When $Q^{2}<<m^{2}$, we get
$$
\Lambda_{\mu}(p',p)=\frac{\alpha}{4\pi}(\frac{11}{2}-\ln\frac{m^{2}}{\mu_{1}^{2}}
+4\ln\eta)\gamma_{\mu}+i\frac{\alpha}{4\pi}
\frac{q^{\nu}}{m}\sigma_{\mu\nu}-\frac{\alpha}{4\pi}(
\frac{1}{6}+\frac{4}{3}\ln\eta)\frac{q^{2}}{m^{2}}\gamma_{\mu}.
$$
It means that the interaction of the electron with the external potential is
modified
\begin{equation}
-e\gamma_{\mu}\rightarrow -e[\gamma_{\mu}+\Lambda_{\mu}(p',p)].
\end{equation}
Besides the important term
$i\frac{\alpha}{4\pi}\frac{q^{\nu}}{m}\sigma_{\mu\nu}$ in $\Lambda_\mu(p',p)$
which emerges as
the anomalous magnetic moment of electron, the charge modification here
is expressed by a renormalization factor $Z_1$:
\begin{equation}
Z_{1}^{-1}=1+\frac{\alpha}{4\pi}\{[2-\ln\frac{m^2}{\mu_{1}^2}-\frac{1}{2w}F(w)]
+(4\ln\eta+5)\frac{2w m^{2}}{Q^{2}}F(w)+(2\ln\eta+\frac{7}{4})2wF(w)\}.
\end{equation}
The infrared term ($\sim\ln\eta$) is ascribed to
the bremsstrahlung of soft photons [20,22] and can be taken care by KLN
theorem [23]. We will fix $\mu_{1}$ and $\eta$ below.

\vskip 24pt
\noindent
{\bf 4. Beta function at one-loop level in QED}\\
\baselineskip 24pt
\indent Adding all three FDI's at one loop level to the tree diagram, we define
the renormalized charge as usual [2, 20-22]:
\begin{equation}
e_{\rm R}=\frac{Z_{2}}{Z_{1}}Z_{3}^{1/2}e.
\end{equation}
But the Ward-Takahashi Identity (WTI) implies that [20-22]
\begin{equation}
Z_{1}=Z_{2}.
\end{equation}
Therefore
\begin{equation}
\alpha_{\rm R}\equiv\frac{e_{\rm R}^{2}}{4\pi}=Z_{3}\alpha.
\end{equation}
Then set $p^{2}=m^{2}$ in $Z_{2}$ and $Q^{2}=0$ in $Z_{1}$ with $\mu_{1}=
\mu_{2}$, yielding
\begin{equation}
\ln\eta=-\frac{5}{8}.
\end{equation}
For any value of $Q$, the renormalized charge reads
from Eqs. (20)---(22):
\begin{equation}
e_{\rm R}(Q)=e\{1+\frac{\alpha}{\pi}\int_{0}^{1}dx[(x-x^{2})\ln
\frac{Q^{2}(x-x^{2})+m^{2}}{\mu_{3}^{2}}]\}
\end{equation}
\begin{equation}
e_{\rm R}(Q)\sim
e\{1+\frac{\alpha}{2\pi}[\frac{1}{3}\ln\frac{m^{2}}{\mu_{3}^{2}}+
\frac{1}{15}\frac{Q^{2}}{m^{2}}]\} \hskip 0.2in (Q^{2}<<m^{2}).
\end{equation}
The observed charge is defined at $Q^{2}\rightarrow 0$ (Thomson
scattering) limit:
\begin{equation}
e_{\rm obs}=e_{\rm R}|_{Q=0}=e
\end{equation}
which dictates that
\begin{equation}
\mu_{3}=m.
\end{equation}
We see that $e^{2}_{R}(Q)$ increases with $Q^{2}$. For
discussing the running of $\alpha_{\rm R}$ with $Q^{2}$, we define the
Beta function:
\begin{equation}
\beta(\alpha, Q)\equiv Q\frac{\partial}{\partial Q}\alpha_{\rm R}(Q)
\end{equation}
From Eq. (39), one finds:
\begin{equation}
\beta(\alpha,Q)=\frac{2\alpha^{2}}{3\pi}-
\frac{4\alpha^{2}m^{2}}{\pi
Q^{2}}\{1+\frac{2m^{2}}{\sqrt{Q^{4}+4Q^{2}m^{2}}}
\ln\frac{\sqrt{Q^{4}+4Q^{2}m^{2}}-Q^{2}}
{\sqrt{Q^{4}+4Q^{2}m^{2}}+Q^{2}}\}
\end{equation}
\begin{equation}
\beta(\alpha, Q)\simeq\frac{2\alpha^{2}}{15\pi}
\frac{Q^{2}}{m^{2}}, \hskip 0.2in (\frac{Q^{2}}{4m^{2}}<<1)
\end{equation}
\begin{equation}
\beta(\alpha, Q)\simeq\frac{2\alpha^{2}}{3\pi}-
\frac{4\alpha^{2}m^{2}}{\pi Q^{2}}, \hskip 0.2in
(\frac{4m^{2}}{Q^{2}}<<1)
\end{equation}
which leads to the well known result $\beta(\alpha)=\frac{2\alpha^{2}}
{3\pi}$ at one loop level at $Q^{2} \rightarrow \infty$.

\vskip 24pt
\noindent
{\bf 5. The RGE in QED with contributions from 9 kinds of fermions with masses}\\
\baselineskip 24pt
\indent Usually, the GRE in QED is obtained by set $Q\longrightarrow\infty$
and $\alpha\longrightarrow\alpha_R(Q)$ in the right hand side of Eq. (43),
\begin{equation}
Q\frac{\partial}{\partial Q}\alpha_R=\frac{2\alpha^2_R}{3\pi}.
\end{equation}
Then after integration, one yields analytically (see Eq. (1)):
\begin{equation}
\alpha_R(Q)=\frac{\alpha}{1-\frac{2\alpha}{3\pi}\ln\frac{Q}{m}}.
\end{equation}
However the renormalization is forced to be made at $Q=m$ so that
\begin{equation}
\alpha_R|_{Q=m}=\alpha.
\end{equation}
\indent We are now in a position to improve the above GRE calculation in three
aspects as indicated at the beginning of this paper.
For constructing a new GRE, we replace the constant $\alpha$ in right
hand side of Eq. (44) by $\alpha_R(Q)$ and add all the contributions from
charged leptons and quarks together, yielding:
\begin{equation}
Q\frac{d}{dQ}\alpha_R(Q)=\sum\limits_i\epsilon_i\left\{\frac{2\alpha^2_R(Q)}{3\pi}
-\frac{4\alpha^2_R(Q)m^2_i}{\pi Q}\left[1+\frac{2m^2_i}{\sqrt{Q^4+4Q^2m^2_i}}
\ln\frac{\sqrt{Q^4+4Q^2m^2_i}-Q^2}{\sqrt{Q^4+4Q^2m^2_i}+Q^2}\right]\right\}
\end{equation}
where
\begin{equation}
\epsilon_i=\left\{
\begin{array}{ll}
1, & i=e,\mu,\tau \\
3\times(\frac{2}{3})^2=\frac{4}{3}, & i=u,c,t \\
3\times(-\frac{1}{3})^2=\frac{1}{3}, & i=d,s,b.
\end{array}
\right.
\end{equation}
\indent Adding up contributions from particles with mass $m_e$, $m_\mu$,
$m_\tau$, $m_c=1.031GeV$, $m_b=4.326GeV$, $m_t=175GeV$ we calculate the
running coupling constant numerically from $\alpha_R(Q=0)=\alpha$ till
\begin{equation}
\alpha_R(Q=m_Z=91.1884GeV)=(131.51)^{-1}
\end{equation}
in comparision with the experimental value [24],
\begin{equation}
\alpha_{exp}(Q=m_Z)=(128.89)^{-1}.
\end{equation}
\indent The remaining discrepancy is ascribed to the contribution of light
quarks $(u,d,s)$ with average mass
\begin{equation}
\bar{m}_q=92MeV, \hskip 0.3in q=u,d,s.
\end{equation}
If we adopt the following values for the mass of light quark:
\begin{equation}
m_u=8MeV, \hskip 0.3in m_d=10MeV, \hskip 0.3in m_s=200MeV
\end{equation}
which are not far from the ratios found by Yan {\it et al.} [25] via the
analysis of mass spectuum of mesons, then the fit will be rather good. See
Fig. 1.

\vskip 48pt
\noindent
{\bf III. RGE of RCC in QCD}

\vskip 24pt
\noindent
\baselineskip 24pt
{\bf 1. Self-energy of quark with mass $m_i$}\\
\indent For convenience, we use the notation and diagram in Ref. [2] at
one-loop level. Then the self-energy of quark with momentum $p$ reads
\begin{equation}
\Sigma_i(p)=-i(A_i+B_i\not{p}).
\end{equation}
The similar procedure as in previous section leads to the renormalization
constant for wave function:
\begin{equation}
Z_{2i}=(1-B_i)^{-1}\approx 1+B_i(p,m_i)
\end{equation}
\begin{equation}
Z_{2i}=1+\frac{\alpha_s}{4\pi}T^aT^a\{\ln\frac{m_i^2}{\mu_{2i}^2}-3
-\frac{(m_i^2-p^2)}{p^2}[1
+\frac{(m_i^2+p^2)}{p^2}\ln\frac{(m_i^2-p^2)}{m_i^2}]\}
\end{equation}
where $\alpha_s=\frac{g_s^2}{4\pi}$ is the strong coupling constant,
$T^aT^a=\frac{4}{3}$, and $\mu_{2i}$ is an arbitrary constant like that in
Eq. (7).

\vskip 24pt
\noindent
\baselineskip 24pt
{\bf 2. Self-energy of gluon}\\
\indent The combination of contributions from the gluon loop and the
Faddeev-Popov ghost field leads to
\begin{equation}
\Pi^g_{\mu\nu,ab}(q)=\frac{i\alpha_s}{4\pi}\delta_{ab}C_A\frac{5}{3}
(g_{\mu\nu}Q^2+q_\mu q_\nu)\ln\frac{Q^2}{\mu^2_3}
\end{equation}
where $Q^2=-q^2>0$, $C_A=3$, and $\mu_3$ being an another arbitrary constant
(See Eq. (20)).\\
\indent The third contribution is coming from quark loop with mass $m_i$
$(i=u,d,s,c,b,t)$:
\begin{equation}
\Pi^{q_i}_{\mu\nu,ab}(q)=\frac{i\alpha_s}{\pi}\delta_{ab}(q_\mu q_\nu
-g_{\mu\nu}q^2)\int^1_0dx(x^2-x)\ln\frac{m_i^2+q^2(x^2-x)}{\mu_3^2}
\end{equation}
(the quark notation $q_i$ should not be confused with the momentum transfer
$q$).\\
\indent Combination of Eq. (59) with (60) induces the change of $\alpha_{s}$:
$$
\alpha_s\longrightarrow Z_3\alpha_s
$$
with
\begin{equation}
Z_3=1+\frac{\alpha_s}{4\pi}[-\frac{5}{3}C_A\ln\frac{Q^2}{\mu_3^2}
+\sum\limits^t_{i=u}4\int^1_0dx(x-x^2)\ln\frac{m_i^2+Q^2(x-x^2)}{\mu_3^2}].
\end{equation}

\vskip 24pt
\noindent
\baselineskip 24pt
{\bf 3. Vertex functions in QCD}\\
\indent There are two kinds of vertex function for one species of quark with
mass $m_i$ at one-loop level in QCD, $\bar{\Gamma}_{\mu i}^{(1)}(q)$ and
$\bar{\Gamma}_{\mu i}^{(2)}(q)$ (see Ref. [2]):
\begin{eqnarray}
\bar{\Gamma}^{(1)}_{\mu i}(q) & = & \frac{\alpha_s}{4\pi}(\frac{C_A}{2}
-T^aT^a)\{[\ln\frac{m_i^2}{\mu_1^2}-\frac{3}{2}+\frac{1}{2\omega_i}F(\omega_i)]
\gamma_\mu \nonumber \\
 & & -(4\ln\eta+5)\frac{2\omega_i}{Q^2}F(\omega_i)m^2_i\gamma_\mu
-2(2\ln\eta+\frac{7}{4})\omega_iF(\omega_i)\gamma_\mu\}
\end{eqnarray}
\begin{eqnarray}
\bar{\Gamma}^{(2)}_{\mu i}(q) & = & \frac{\alpha_s}{4\pi}\frac{C_A}{2}
\int^1_0du\int^{+u}_{-u}dv\{-3\gamma_\mu(\ln\frac{M_i^2}{\mu_1^2}+\frac{1}{2})
\nonumber \\
 & & +\gamma_\mu\frac{[2u(1-u)m_i^2+\frac{q^2}{2}(u^2-u-v^2)]}{2M_i^2}\}
\end{eqnarray}
where $\mu_1$ $(\eta)$ is an arbitrary constant introduced for dealing with
the ultraviolet (infrared) divergence (see Eqs. (23) --- (34)),
\begin{equation}
\omega_i=\frac{1}{\sqrt{1+4m_i^2/Q^2}}, \hskip 0.3in
F(\omega_i)=ln\frac{1+\omega_i}{1-\omega_i}
\end{equation}
\begin{equation}
M_i^2=m_i^2(1-u)^2+\frac{Q^2}{4}(u^2-v^2).
\end{equation}
Here the new renormalization method has been used and two terms related
to the anomalous magnetic moment of quarks have been omitted. The two
Feynman diagrams give the correction of vertex function at one-loop level
\begin{equation}
-ig_sT^a\gamma_\mu\longrightarrow -ig_sT^a(\gamma_\mu
+\bar{\Gamma}^{(1)}_{\mu i}+\bar{\Gamma}^{(2)}_{\mu i})
=-ig_sT^a\gamma_\mu/Z_{1i}.
\end{equation}
Then,
\begin{eqnarray}
Z^{-1}_{1i} & = & 1+\frac{\alpha_s}{4\pi}(\frac{C_A}{2}-T^aT^a)
\{\ln\frac{m_i^2}{\mu_1^2}-\frac{3}{2}+\frac{1}{2\omega_i}F(\omega_i)
-(4\ln\eta+5)\frac{2m_i^2}{Q^2}\omega_iF(\omega_i) \nonumber \\
 & & -2(2\ln\eta+\frac{7}{4})\omega_iF(\omega_i)\}+\frac{\alpha_s}{4\pi}
\frac{C_A}{2}\int^1_0du\int^{+u}_{-u}dv\{-3\ln\frac{M_i^2}{\mu_1^2}-\frac{3}{2}
\nonumber \\
 & & +\frac{u(1-u)m_i^2+\frac{Q^2}{4}(u-u^2+v^2)}{m_i^2(1-u)^2
+\frac{Q^2}{4}(u^2-v^2)}\}.
\end{eqnarray}

\vskip 24pt
\noindent
\baselineskip 24pt
{\bf 4. Beta function at one-loop level in QCD}\\
\indent Combining all of the above one-loop Feynman diagrams and considering
$p=\frac{q}{2}$ in $Z_{2i}$, the strong coupling constant $\alpha_s$ is
modified to
\begin{equation}
\alpha_s\longrightarrow \alpha_{si}(Q,m_i)
=\frac{Z^2_{2i}Z_3}{Z^2_{1i}}\alpha_s.
\end{equation}
For discussing the running of $\alpha_{si}(Q,m_i)$ with $Q^2$, we define
the $\beta$-function
\begin{eqnarray}
\beta_i(Q,m_i) & = & Q\frac{\partial}{\partial Q}\alpha_{si}(Q,m_i)
=2Q^2\frac{\partial}{\partial Q^2}\alpha_{si}(Q,m_i) \nonumber \\
 & = & 2Q^2\alpha_s(\frac{\partial}{\partial Q^2}Z^2_{2i}
+\frac{\partial}{\partial Q^2}Z_3+\frac{\partial}{\partial Q^2}Z^{-2}_{1i}).
\end{eqnarray}
By denoting
$$
\frac{\partial}{\partial Q^2}Z^2_{2i}=\frac{\alpha_s}{4\pi Q^2}B_{2i}(Q,m_i)
$$
\begin{equation}
\frac{\partial}{\partial Q^2}Z_3
=\frac{\alpha_s}{4\pi Q^2}B_3(Q,m_u,\cdots,m_t)
\end{equation}
$$
\frac{\partial}{\partial Q^2}Z^{-2}_{1i}
=\frac{\alpha_s}{4\pi Q^2}B_{1i}(Q,m_i),
$$
we get
\begin{equation}
\beta_i(Q,m_i)=\frac{\alpha^2_s}{2\pi}(B_{1i}+B_{2i}+B_3).
\end{equation}

\vskip 24pt
\noindent
\baselineskip 24pt
{\bf 5. RGE for quark $q_i$ with mass $m_i$ in QCD}\\
\indent The RGE is established by simply substituting the $\alpha_s$ by
$\alpha_{si}(Q,m_i)$ at the right side, yielding
\begin{equation}
Q\frac{\partial}{\partial Q}\alpha_{si}(Q,m_i)
=\frac{1}{2\pi}(B_{1i}+B_{2i}+B_3)\alpha^2_{si}(Q,m_i).
\end{equation}

\vskip 48pt
\noindent
{\bf IV. Numerical calculation of RGE in QCD}

\vskip 24pt
\baselineskip 24pt
\indent Obviously, Eq. (72) can only be integrated numerically for one species
of quark with mass $m_i$. We adopt the experimental data $Q=m_Z=91.1884GeV$,
$\alpha_{si}=0.118$ [26,27] as the initial value of integration. Then,
$\alpha_{si}(Q,m_i)$ becomes
\begin{equation}
\alpha_{si}(Q,m_i)=\frac{1}{\frac{1}{0.118}
+\frac{1}{2\pi}\int^{91188.4}_Q(B_{1i}+B_{2i}+B_3)\frac{1}{Q}dQ}
\end{equation}
where
\begin{eqnarray}
B_{1i}(Q,m_i) & = & \frac{1}{3}(\frac{m_i^2}{Q^2}\omega_iF(\omega_i)
+\frac{m_i^2}{Q^2}(1-\frac{4m_i^2}{Q^2})\omega_i^3F(\omega_i)
+(\frac{1}{2}-\frac{2m_i^2}{Q^2})\omega_i^2+\frac{1}{2}) \nonumber \\
 & & -9+3\int^1_0duG_i(u,Q)
\end{eqnarray}
\begin{equation}
G_i(u,Q)=\frac{4m_i^2}{Q^2}(1-u)(u-2u^2-\frac{1}{2\xi_i})
\ln\frac{\xi_i+u}{\xi_i-u}+\frac{1}{\xi_i^2}(u^2+\frac{4m_i^2}{Q^2}u(1-u))
\end{equation}
\begin{equation}
\xi_i=\sqrt{\frac{4m_i^2}{Q^2}(1-u)^2+u^2}, \hskip 0.3in
\omega_i=\frac{1}{\sqrt{1+4m_i^2/Q^2}}, \hskip 0.3in
F(\omega_i)=\ln\frac{1+\omega_i}{1-\omega_i},
\end{equation}
\begin{equation}
B_{2i}(Q,m_i)=\frac{8}{3}(1+\frac{8m_i^2}{Q^2}(-1+\frac{4m_i^2}{Q^2}
\ln(1+\frac{Q^2}{m_i^2})))
\end{equation}
\begin{equation}
B_3(Q,m_u,\cdots,m_t)=-1-\sum\limits^t_{i=u}(\frac{4m_i^2}{Q^2}
-\frac{8m_i^4}{Q^4}\omega_iF(\omega_i)).
\end{equation}
The results are shown in Figures 2 and 3.

\vskip 48pt
\noindent
{\bf V. Summary and discussion}

\vskip 24pt
\baselineskip 24pt
\indent 1. Let us first check the zero mass limit of above equations for
returning to the familiar result Eq. (2). For the purpose we look directly at
the $Z_i$ in the limit $m_i/Q\longrightarrow 0$, yielding
\begin{equation}
\left\{
\begin{array}{l}
Z^{-1}_{1}=1-\frac{\alpha}{4\pi}(C_A+T^aT^a)\ln\frac{Q^2}{\mu^2} \\
Z_2=1+\frac{\alpha}{4\pi}T^aT^a\ln\frac{Q^2}{\mu^2} \\
Z^{1/2}_3=1+\frac{\alpha}{8\pi}(\frac{4}{3}C_f-\frac{5}{3}C_A)
\ln\frac{Q^2}{\mu}
\end{array}
\right.
\end{equation}
where we have chosen $\ln\eta=-1$ with another constants
$\mu_1=\mu_2=\mu_3=\mu$. This recipe amounts to define the value of
$\alpha_s$ at high $Q$ limits.\\
\indent Substituting Eq. (79) into Eq. (71), we obtain
\begin{equation}
\beta(Q)=-\frac{\alpha^2}{2\pi}\beta_0, \hskip 0.3in
\beta_0=\frac{11}{3}C_A-\frac{2}{3}n_f, \hskip 0.3in C_A=3.
\end{equation}
Then the RGE reads
\begin{equation}
Q\frac{\partial}{\partial Q}\alpha_R(Q)=-\frac{1}{2\pi}\beta_0\alpha^2_R(Q)
\end{equation}
with its solution precisely giving Eq. (2).

\vskip 24pt
\baselineskip 24pt
\indent 2. Alternatively, we manage to keep the quark mass in all $B_i$ to
get the RGE (72) before setting the limit $m_i\longrightarrow 0$:
$$
\left.
\begin{array}{l}
B_2\longrightarrow 2T^aT^a \\
B_3\longrightarrow -\frac{5}{3}C_A+\frac{2}{3}n_f \\
B_1\longrightarrow 2(\frac{C_A}{2}-T^aT^a)-2C_A.
\end{array}
\right.
$$
Thus, in the limit $m_i\longrightarrow 0$,
\begin{equation}
B_1+B_2+B_3\longrightarrow\frac{2}{3}n_f-\frac{8}{3}C_A=-\beta_0'.
\end{equation}
It is interesting to compare (82) with (80), showing that
\begin{equation}
\beta_0-\beta_0'=C_A
\end{equation}
which is stemming from the different order of taking limit: either
$m_i\longrightarrow 0$ before the derivative $\frac{\partial}{\partial Q^2}$
or vice versa.

\vskip 24pt
\baselineskip 24pt
\indent 3. But the zero mass limit is certainly not a good one as discussed
in the introduction. And this is why one usually had to take $n_f=3$ in
$\beta_0$. The mass of $c$ or $b$ quark is too heavy to be neglected.
Therefore, we have calculated seriously the RGE for five quarks
$(u,d,s,c,b)$ with masses except $t$ quark. The latter is too heavy to be created
explicitly in the energy region considered. Notice that, however, the
contribution of $t$ quark is still existing in the function $B_3$, Eq. (78).

\vskip 24pt
\baselineskip 24pt
\indent 4. The prominent feature of our RGE calculation is the following:\\
\indent (a) The RCC $\alpha_{si}(Q,m_i)$ has a flavor dependence, i.e., it is
different for different quark with different $m_i$.\\
\indent (b) The value of $\alpha_{si}(Q,m_i)$ increases from normalized value
0.118 at $Q=M_Z=91.1884GeV$ with the decrease of $Q$ until a maximum
$\alpha^{max}_{si}$ is reached at $Q=\Lambda_i$. The smaller the $m_i$ is, the
smaller the $\Lambda_i$ is and the higher the value of $\alpha^{max}_{si}$
will be. When $Q\longrightarrow 0$, all $\alpha_{si}$ approach to zero.\\
\indent (c) The value of $\Lambda_i$ could be explaned as the existence of a
critical length scale $L_i$ of $q_i\bar{q}_i$ pair
\begin{equation}
L_i\sim\hbar/\Lambda_i
\end{equation}
while the value $\alpha^{max}_{si}$ may correspond to the excitation energy for
breaking the binding $q_i\bar{q}_i$ pair, i.e., the threshold energy scale
against its dissociation into two bosons:
\begin{equation}
E^{thr}_i\sim\alpha^{max}_{si}/L_i\sim\alpha^{max}_{si}\Lambda_i/\hbar.
\end{equation}
The numerical estimation of these values is listed at the table 1. It is
interesting to see that $E^{thr}_i$ for $u$, $d$ quarks is of the order of
$\pi$ meson while that for $c$ or $b$ quark could be compared with the
$D^+D^-$ or $B^+B^-$ threshold respectively.

\vskip 24pt

\begin{center}
\tabcolsep 0.7cm
\begin{tabular}{|c|c|c|c|c|c|}
\hline
& u & d & s & c & b \\ \hline
m$_i$c$^2$(MeV)  & 8 & 10 & 200 & 1031 & 4326 \\ \hline
$\Lambda _i$(MeV) & 18.4 & 18.4 & 290 & 1640 & 7040 \\ \hline
$\alpha _{si}^{\max }$ & 12.43 & 9.368 & 0.3027 & 0.2038 & 0.1610 \\ \hline
L$_i$(fm) & 10.73 & 10.73 &  0.6809  &  0.1204  &  0.02805  \\ \hline
E$_i^{thr}$(MeV) &  228.7  &  172.4  & 87.77 & 334.3 & 1133 \\ \hline
\end{tabular}

\vskip 24pt

{\bf Table 1}
\end{center}

\vskip 24pt
\baselineskip 24pt
\noindent
{\bf Acknowledgements}\\
\indent We thank Prof. Xiao-tong Song and Dr. Ji-feng Yang for
discussions. This work was supported in part by the NSF of China.

\vskip 48pt
\noindent
{\bf References}

\vskip 24pt
\baselineskip 24pt
\noindent
\mbox{\hspace{5pt}}[1] I. Aitchison and A. Hey, Gauge Theories in Particle
Physics (Adam Hilger LTD, Bristol, 1982), pp 284, 287.\\
\noindent
\mbox{\hspace{5pt}}[2] R.D. Field, Application of Perturbative QCD
(Addison-Wesley Publishing Company, 1989).\\
\noindent
\mbox{\hspace{5pt}}[3] H. Epstein and V. Glaser, Ann. Inst. Hemi. Poincare,
19 (1973) 211.\\
\noindent
\mbox{\hspace{5pt}}[4] J. Collins, Renormalization (Cambridge University
Press, 1984).\\
\noindent
\mbox{\hspace{5pt}}[5] G. Scharf, Finite Electrodynamics (Springer-Verlag,
Berlin, 1989).\\
\noindent
\mbox{\hspace{5pt}}[6] J. Glimm and A. Jaffe, Collective Papers, Vol.2
(1985).\\
\noindent
\mbox{\hspace{5pt}}[7] M. D\"{u}tch, F. Krahe and G. Scharf, Phys. Lett. B258
(1991), 457.\\
\noindent
\mbox{\hspace{5pt}}[8] D.Z. Freedmann, K. Johnson and J.I. Lattore, Nucl.
Phys. B371 (1992), 353.\\
\noindent
\mbox{\hspace{5pt}}[9] P.E. Haagensen and J.I. Latorre, Phys. Lett. B 283,
(1992) 293.\\
\noindent
[10] G. Dunne and N. Rius, ibid 293 (1992) 367.\\
\noindent
[11] V.A. Smirnov, Nucl. Phys. B 427 (1994) 325.\\
\noindent
[12] Ji-feng Yang, Thesis for PhD. (Fudan University, 1994); Preprint,
(1997) hep-th/9708104; Ji-feng Yang and G-j Ni, Acta Physica Sinica (Overseas
Edition), 4 (1961) 88.\\
\noindent
[13] Guang-Jiong Ni and Su-qing Chen, Acta Physics Sinica
(Overseas Edition) 7 (1998) 401; Internet, hep-th/9708155.\\
\noindent
[14] G-j Ni, S-y Lou, W-f Lu and J-f Yang, Science in China (Series A) 41,
(1998) 1206; Internet, hep-ph/9801264.\\
\noindent
[15] G-j Ni, S-q Chen, W-f Lu, J-f Yang and S-y lou,
$\ll$Frontiers in Quantum Field Theory$\gg$
Edit: C-Z Zha and K. Wu (World Scientific, 1998), pp 169-176.\\
\noindent
[16] G-j Ni and H. Wang, $\ll$Physics Since Parity Symmetry
Breaking$\gg$ Edit: F. Wang (World Scientific,
1998), pp 436-442; Preprint,
Internet, hep-th/9708457.\\
\noindent
[17] G-j Ni, Kexue (Science) 50(3) (1998) 36-40; Internet,
quant-ph/9806009, to be published in a book $\ll$Photon: Old
problems in light of New ideas$\gg$ Edit: V. Dvoeglazov (Nova
Publisher, 1999).\\
\noindent
[18] G-j Ni and H. Wang, Journal of Fudan University (Natural Science) 37(3)
(1998) 304-305.\\
\noindent
[19] J.J. Sakurai, Advanced Quantum Mechanics (Addison-Wesley Publishing
company, 1967).\\
\noindent
[20] J.D. Bjorken and S.D. Drell, Relativistic Quantum Mechanics (McGraw-Hill
Book Company, 1964).\\
\noindent
[21] P. Ramond, Field Theory: A Modern Primer (The Benjamin/Cummings Publishing
Company, 1981).\\
\noindent
[22] C. Itzykson and J.B. Zuber, Quantum Field Theory (McGraw-Hill Inc., 1980).\\
\noindent
[23] T.D. Lee, Particle Physics and Introduction to Field Theory (Harwood
Academic Publishers, 1981).\\
\noindent
[24] H. Burkhardt and B. Pietrzyk, Phys. Lett. B356 (1995) 398.\\
\noindent
[25] D. Gao, B. Li and M. Yan, Phys. Rev. D 56 (1997) 4115.\\
\noindent
[26] Maria Girone and Matthias Neubert, Phys. Rev. Lett. 76 (1996) 3061.\\
\noindent
[27] Michael Schmelling, hep-ex/9701002.

\vskip 48pt
\normalsize
\baselineskip 24pt
\noindent
{\bf Figure Caption}

\vskip 24pt
\baselineskip 24pt
\normalsize
\noindent
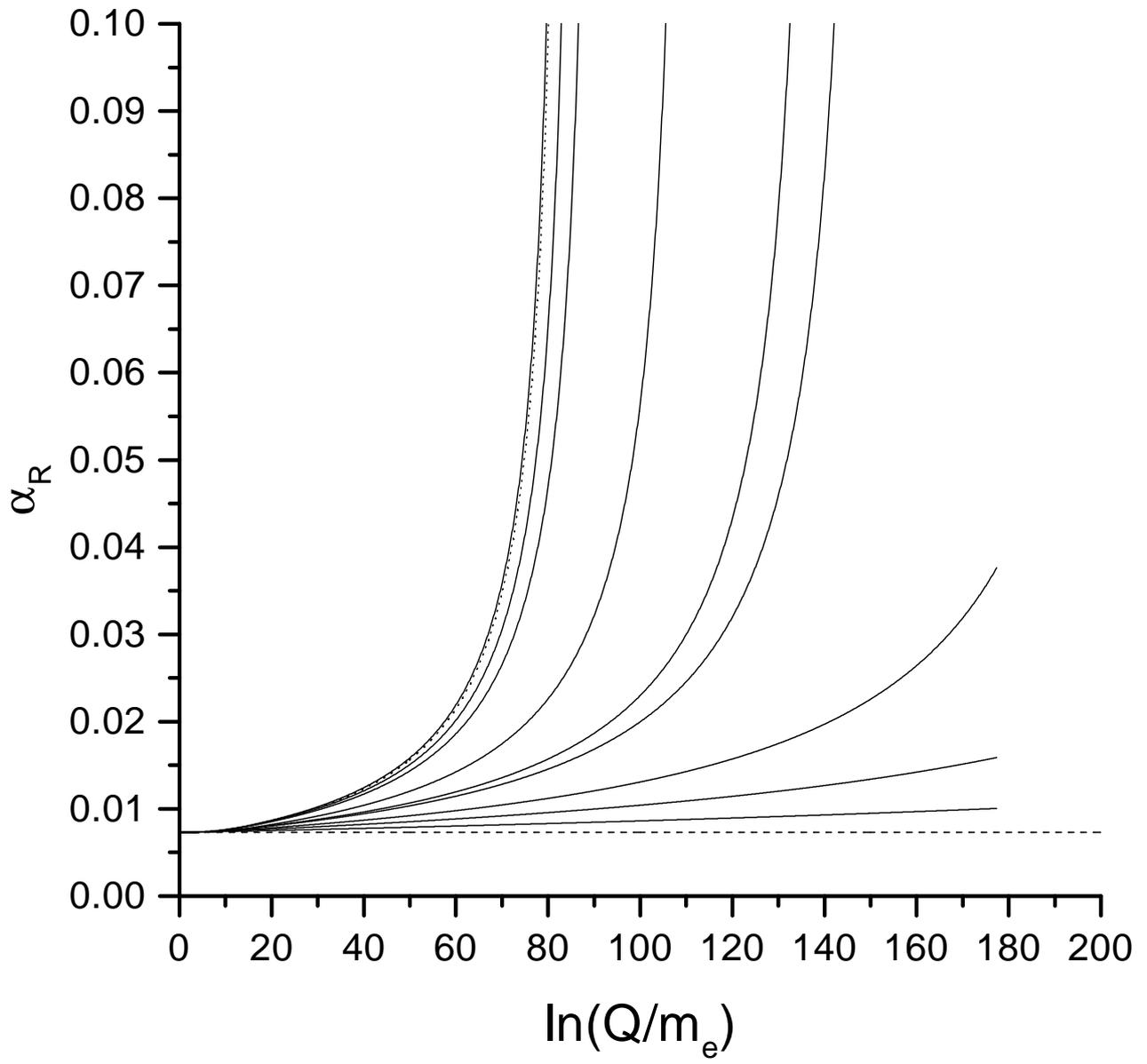
{\bf Figure 1:}\\
\baselineskip 24pt
\indent The nine curves (see from the lowest) represent respectively the
contributions to the running electromagnetic coupling constant from\\
\indent (1) electron $e$ only,\\
\indent (2) $e$ and muon ($\mu$) only,\\
\indent (3) all charged leptons $e$, $\mu$ and $\tau$ only,\\
\indent (4) $e$, $\mu$, $\tau$ and $c$ quark only,\\
\indent (5) $e$, $\mu$, $\tau$, $c$ and $b$ quark only,\\
\indent (6) $e$, $\mu$, $\tau$, $c$, $b$ and $t$ quark only,\\
\indent (7) $e$, $\mu$, $\tau$, $c$, $b$, $t$ and $u$ quark only,\\
\indent (8) $e$, $\mu$, $\tau$, $c$, $b$, $t$, $u$ and $d$ quark only,\\
\indent (9) all charged leptons and quarks.\\
The last curve is actually coinciding with the experimental curve denoted by
dot line which can also be fitted by assuming three light quarks ($u$, $d$,
$s$) having average mass $92Mev/c^2$.

\vskip 24pt
\baselineskip 24pt
\normalsize
\noindent
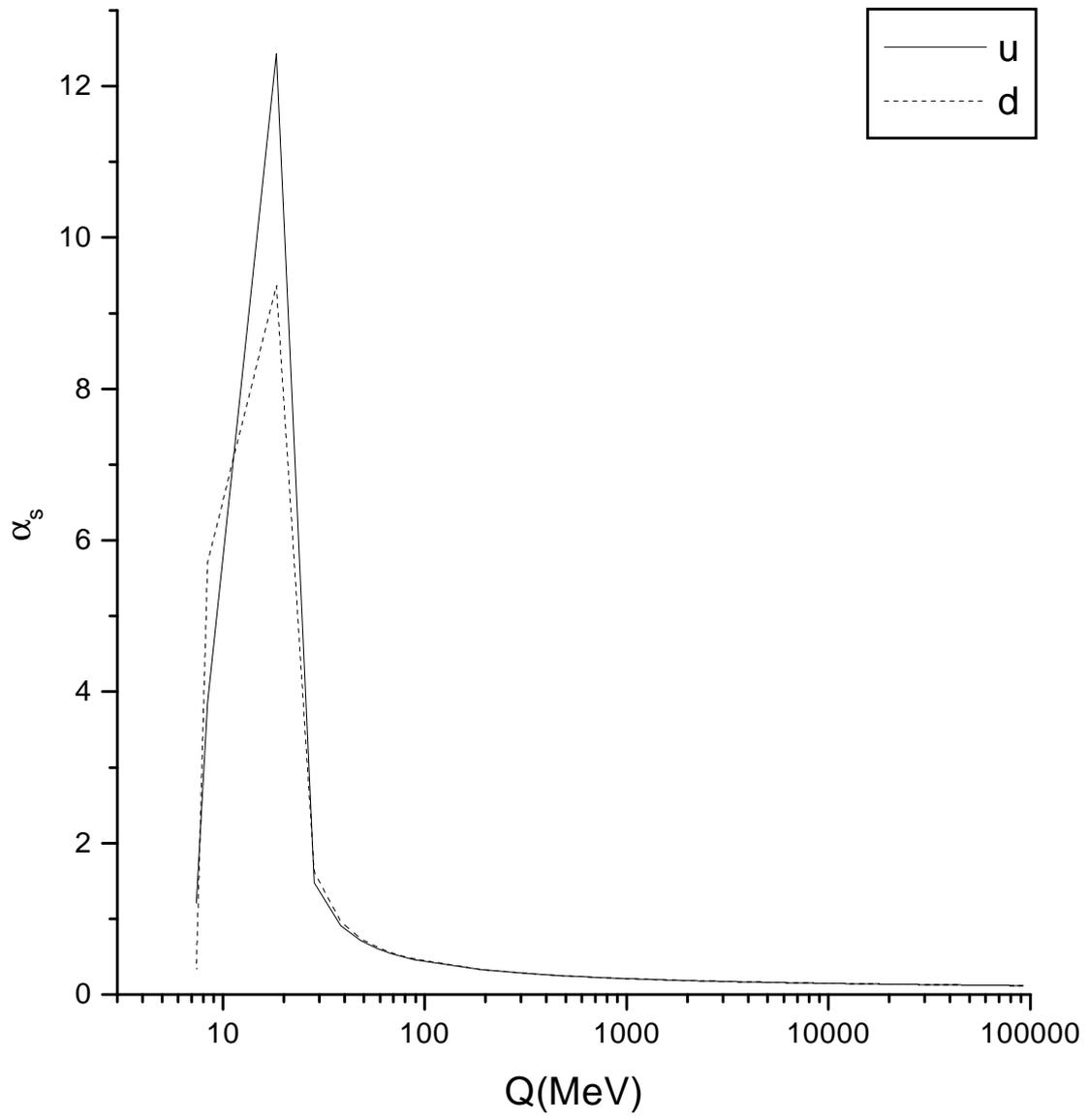
{\bf Figure 2:}\\
\baselineskip 24pt
\indent The running strong coupling constant curves for $u$ and $d$ quarks.

\vskip 24pt
\baselineskip 24pt
\normalsize
\noindent
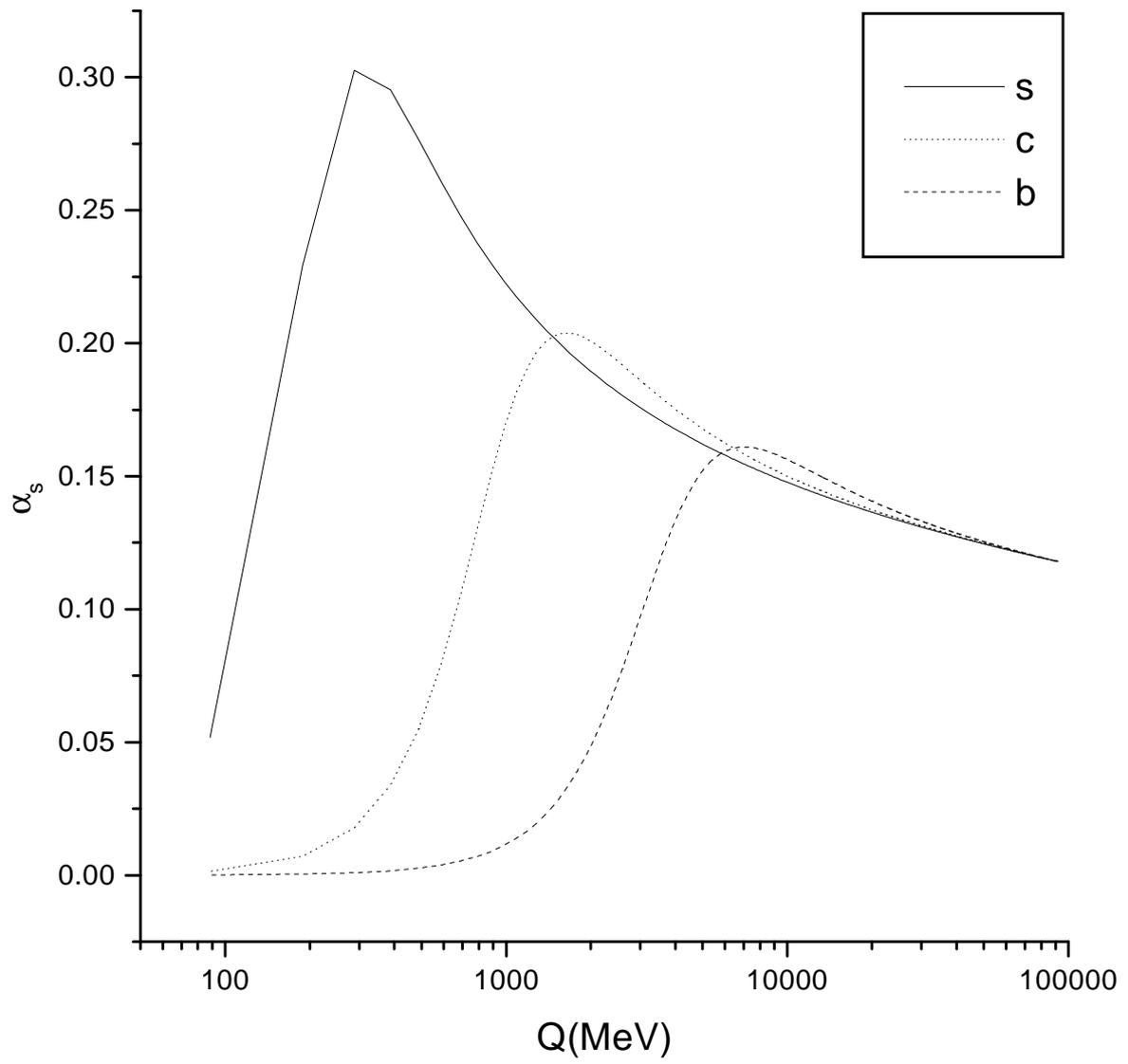
{\bf Figure 3:}\\
\baselineskip 24pt
\indent The running strong coupling constant curves for $s$, $c$ and $b$
quarks.
\end{document}